\documentclass{article}

\usepackage{abstract}

\usepackage{amsmath}

\usepackage{authblk}

\usepackage[style=numeric-comp, sorting=none, giveninits=true, date=year]{biblatex}
\DeclareNameAlias{author}{last-first}
\renewbibmacro{in:}{}
\DeclareFieldFormat{pages}{#1}
\usepackage[labelfont=bf]{caption}

\usepackage[margin=1in]{geometry}

\usepackage{graphicx}
\graphicspath{ {./images/} }

\usepackage[hyperfootnotes=false]{hyperref}
\hypersetup{
    colorlinks=true,
    linkcolor=blue,
    citecolor=blue,
    filecolor=magenta,
    urlcolor=blue,
}

\usepackage{lineno}

\usepackage{siunitx}

\usepackage{url}

\begin{document}

\title{\textbf{Depth-dependent resolution quantification in 3D fluorescence microscopy}}

\author[1]{Neil Wright}
\author[1,*]{Christopher J Rowlands}

\affil[1]{Department of Bioengineering, Imperial College London, London, UK}
\affil[*]{\href{mailto:c.rowlands@imperial.ac.uk}{c.rowlands@imperial.ac.uk}}

\date{}

\maketitle

\begin{abstract}

  A method is presented to quantify resolution as a function of depth in features of morphologically complex 3D samples. Applying the method to the brain of Drosophila, resolution is measured at increasing depth throughout the central brain region. The results quantify improvements in image quality when using two-photon microscopy compared to confocal. It is also demonstrated how resolution improvements through tuning a single parameter, laser power, can be measured objectively. Since the metric is interpretable as the average resolution within a feature, it is suitable for comparing results across optical systems, and can be used to inform the design of biological experiments requiring resolution of structures at a specific scale.

\end{abstract}

\section*{Introduction}

Quantifying image quality is important for both experiment design and the development of optical instruments. In biology, it may be necessary to resolve structures at a certain level of detail; users may not appreciate that features which can be readily observed in one part of the sample may be unresolvable elsewhere. Similarly, having an objective metric of quality in microscopy allows comparison and improvement of instruments under various adverse imaging conditions, as opposed to the favourable conditions that occur near a tissue surface. Since the final quality is a function of both the sample and optical system as a whole, this should be reflected in any metric.

In 3D samples, image quality is also highly dependent on tissue depth, with quality degrading due to light attenuation and distortion caused by scattering. Previous approaches to quantify this effect have sometimes relied on signal intensity \cite{balu:2009, dong:2000} as a metric. However, intensity lacks an obvious practical interpretation, and may not always correlate with image quality. Another approach is to use a score based on arbitrary units \cite{preusser:2021}, though this also has the same issue of interpretability. In theory, using resolution as a metric directly can solve these issues.

Resolution refers to the minimum distance at which two separate objects are distinguishable. Mathematically, this can be defined based on either spatial frequency contrast or distance. The Modulation Transfer Function (MTF) can be used to characterise the former, while the Rayleigh Criterion is an example of the latter. This states that two Airy discs are resolvable if the centre of one disc lies within or outside the first minimum of the diffraction pattern of the other \cite{rayleigh:1879}.

While specially manufactured test samples can be used to assess the performance of an optical system using either method, natural images are unlikely to contain patterns of known contrast or distance that can be used directly to measure resolution. Estimation methods must therefore be used instead. This can be done by either estimating the MTF \cite{rowlands:2015} or using an approach based on Fourier Ring Correlation (FRC) \cite{banterle:2013}. The latter technique originates in electron microscopy and involves finding the highest spatial frequency in an image distinguishable from noise \cite{saxton:1982}.

However, a problem arises when applying a single measurement to more complex samples where resolution may be non-uniform across the image. This is often the case in 3D biological specimens where variations in factors such as tissue depth, type, or fluorophore concentration may create large differences in quality within the same optical plane.

Previously, it has been shown that FRC can be used to analyse local resolution by splitting the image into tiles \cite{culley:2018}, an approach recently used to analyse fine features in super resolution images \cite{zhao:2022_a}. Here, by applying this approach to 3D fluorescence microscopy, we show how it can be used to isolate a particular feature within a 3D sample and quantify resolution within that feature as a function of depth.

The method is demonstrated by analysing the central region of the dissected brain of \textit{Drosophila Melanogaster}, a model organism widely used in neurobiology. Due to its 3D morphology, optical sections of the fly brain can contain regions of heterogeneous resolution, with particular differences between the central brain and optic lobes caused by differing levels of light scattering. By calculating the average FRC value within the central brain, we show how to characterise resolution within any arbitrarily shaped region of an image, and by extension quantify the degradation of resolution in a feature in 3D.

Results comparing confocal and two-photon imaging are presented. These align with previous findings on the highly scattering nature of the fly brain compared to mammalian brain, even when two-photon microscopy is used \cite{hsu:2019}. Additionally, we show how to quantify improvements in resolution when a single parameter, laser power, is increased.

The method presented is generally applicable for analysing resolution in features of 3D samples where image quality is non-uniform. In addition, as a metric of resolution, it has a practical interpretation which can be used to inform experimental considerations and compare results across different optical instruments.

\section*{Methods}

\subsection*{Sample preparation}

Green Fluorescent Protein (GFP) was expressed pan-neuronally in \textit{Drosophila} by crossing nSyb-GAL4 (Bloomington Drosophila Stock Center (BDSC) \#68222) with 10XUAS-IVS-mCD8::GFP (BDSC \#32187). Adult male and female flies were anaesthetised by low temperature and brains dissected in phosphate-buffered saline (PBS) using fine forceps (Dumont \#55 and \#5SF) \cite{wu:2006}. Dissected brains were transferred to a PBS-filled glass-bottom 35 mm confocal dish (VWR) coated with poly-D-lysine (approximately \SI{100}{\micro\gram}/mL; Sigma), and oriented dorsal side down.

\subsection*{Confocal and two-photon microscopy}

Confocal and two-photon (2p) images were both taken using the same commercial Leica SP5 inverted microscope at the Facility for Imaging by Light Microscopy (FILM) at Imperial College London. Each sample was used to capture a stack for all modalities. Because initial results found photobleaching to be minimal except for higher power 2p stacks, the order was not randomised and 2p stacks were always captured last.

A 20x 0.7 NA dry objective (Leica) was used for imaging and a photomultiplier tube (PMT) was used to detect fluorescence. Bit depth was set to 16 bits and pixel size set to 94.6 nm (8192$\times$8192 format) with a line scan rate of 400 Hz.

Frames were captured with a step size of \SI{10}{\micro\meter}. Confocal imaging used an Argon laser at 488 nm and pinhole size of 1 Airy Unit. Two-photon imaging used a Mai Tai DeepSee laser (Newport Spectra-Physics) tuned to 920 nm and the pinhole fully opened.

Power was measured using a Thorlabs S170C power sensor for the Argon laser, and S425C for the DeepSee laser.

\subsection*{Image analysis}

\subsubsection*{Mean resolution algorithm}

\begin{figure}[htbp]
        \centering  
        \includegraphics[width=\textwidth]{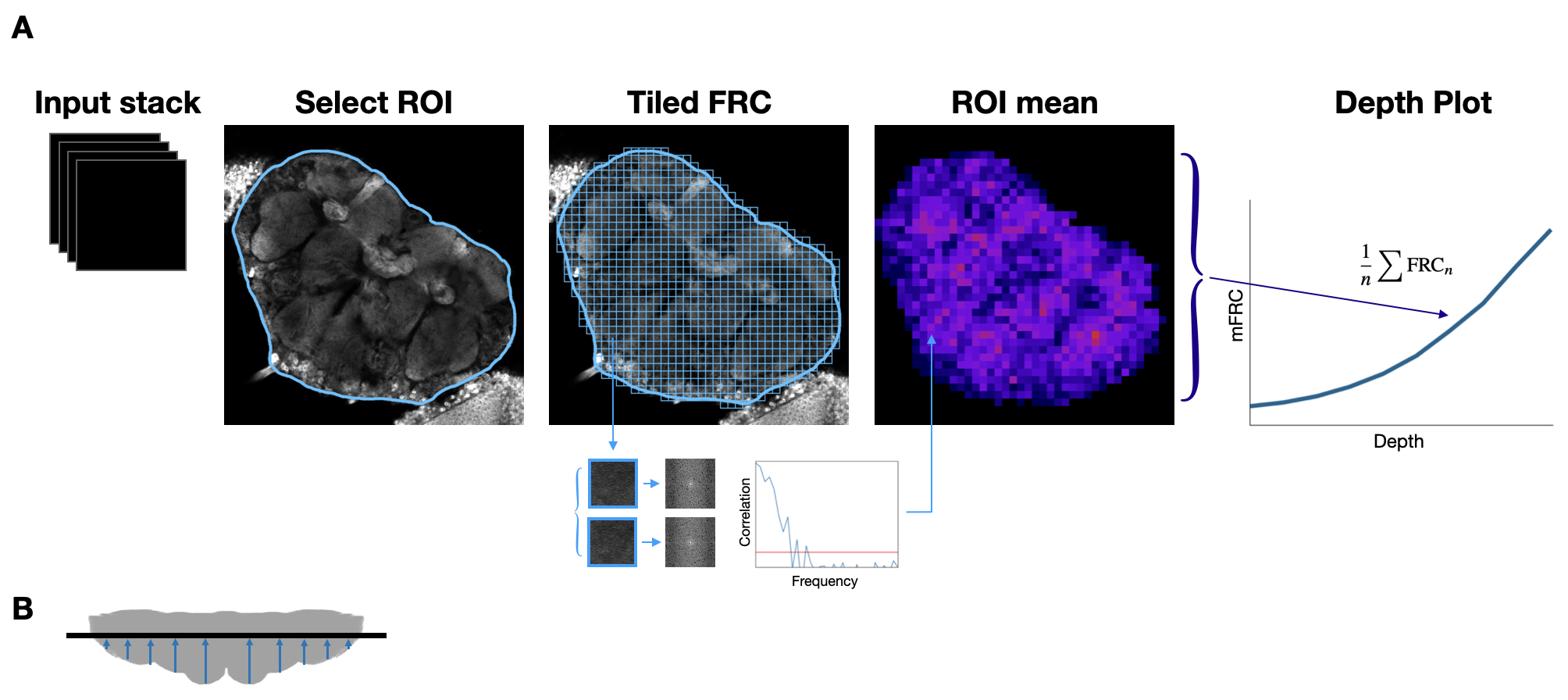}
        \caption{(A) Schematic of the method to calculate Mean FRC (mFRC). First, a region-of-interest (ROI) is drawn over the feature to be analysed in each image in the Z-stack. The image is then divided into small tiles, and the Fourier Ring Correlation (FRC) value for each tile within the ROI is calculated by selecting the highest spatial frequency whose correlation coefficient is greater than the cut-off value. Finally, the mean value for each ROI is computed and plotted as a function of depth. Since FRC represents the minimum resolvable distance, lower values correspond to better image quality. (B) Drosophila brain with the dorsal side facing down. The black line represents an optical plane. Variable tissue depths (blue arrows) result in differing levels of light scattering, contributing to resolution heterogeneity within the image. Drosophila brain based on graphic from virtualflybrain.org \cite{court:2023}.}
        \label{fig:algorithm}
\end{figure}

The algorithm to compute the Mean FRC (mFRC) in a region-of-interest (ROI) of arbitrary shape is illustrated in Figure~\ref{fig:algorithm}. First, ROIs were drawn around the central brain region for each image in the Z-stack and used to generate masks. The image was then split into small (64x64 pixels) tiles. For each tile in the mask, FRC was calculated. The mFRC value for each depth was taken to be the mean of all tiles in the mask for that depth. Finally, mFRC was plotted as a function of depth.

In general, calculating FRC requires two independent images. Here, `single image' FRC was used, whereby each full image is first split into sub-images as described in \cite{koho:2019}. FRC was calculated according to the standard formula of  computing the Pearson correlation coefficient of rings of increasing radius in the Fourier transforms of the two images according to Equation~\ref{eq:pearson}:

\begin{equation}
  \text{FRC}(r) = \frac{ \sum\limits_{i \in R} F_1(r_i) \cdot F_2(r_i)* } { \sqrt{ \sum\limits_{i \in R}|F_1(r_i)|^2 \cdot \sum\limits_{i \in R} |F_2(r_i)|^2 } }
  \label{eq:pearson}
\end{equation}

where $F_1$ is the Fourier transform of the first image and $F_2*$ is the complex conjugate of the Fourier transform of the second image; the numerator is a real number \cite{vanheel:1987}. The inverse of the frequency below which the correlation drops below a fixed 1/7 cut-off is the FRC value \cite{nieuwenhuizen:2013}. All code was written in Java.

\subsubsection*{FRC Colourmaps}

\begin{figure}[htbp]
	\centering
	\includegraphics[width=\textwidth]{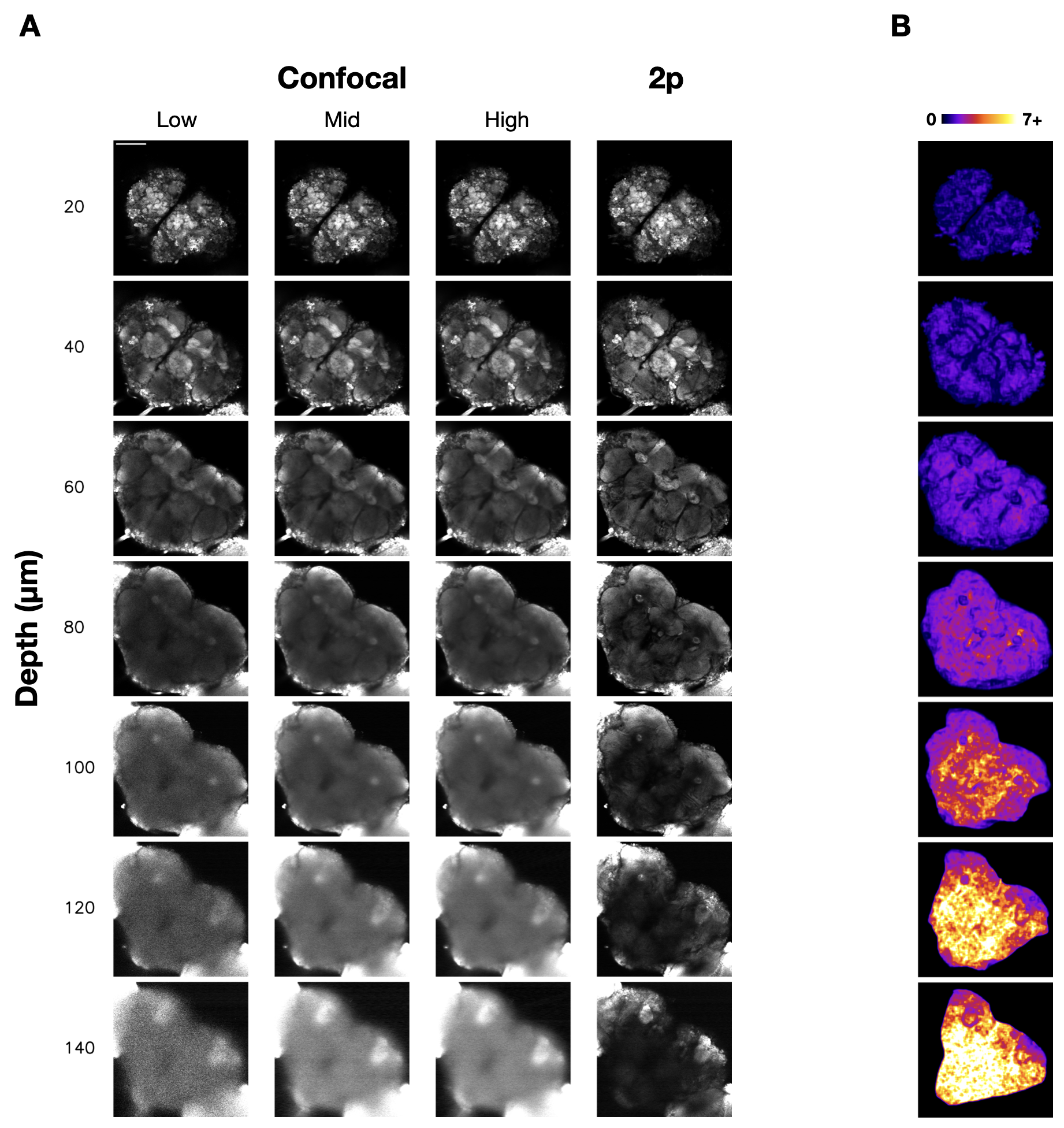}
	\caption{(A) Example images of an nSyb>GFP brain at various depths for each imaging modality. Confocal power levels were \SI{9}{\micro\watt} (low), \SI{100}{\micro\watt} (mid), and \SI{224}{\micro\watt} (high). Two-photon power was 22mW. Each image has been individually optimised for brightness and contrast to highlight image quality. Scale bar = \SI{100}{\micro\meter}. (B) High-detail rolling FRC colourmaps of the central brain ROI, generated from the same two-photon stack shown in A. The scale represents minimum resolvable distance (FRC) from 0 to 7+ \SI{}{\micro\meter}.}
	\label{fig:image_matrix}
      \end{figure}

High-detail FRC colourmaps \cite{zhao:2022_a} were generated to highlight precise variations in resolution across images (Figure~\ref{fig:image_matrix}). The approach to generate colourmaps was similar to the method described above, except that, instead of dividing the image into tiles, a 64x64 pixel block centred on each pixel was scanned across the ROI. The FRC value for each block was converted to a colour value and used to draw a single pixel at the corresponding coordinates in the colourmap. Artifacts were smoothed by applying a Gaussian blur (Radius = 10) using Fiji \cite{schindelin:2012}. It was found that the rolling FRC method produced comparable results to the tiling method in terms of mean value within the ROI, but at the cost of greatly increased computation. Therefore, the tiling method was used when comparing results from different imaging modalities. As a simple example, assuming a tile size of 64 $\times$ 64 pixels, a 1024 $\times$ 1024 pixel square ROI requires 16 $\times$ 16 = 256 FRC calculations for the tiling method, but 1,048,576 FRC calculations for the rolling block method.

\begin{figure}[htbp]
	\centering
	\includegraphics[width=\textwidth]{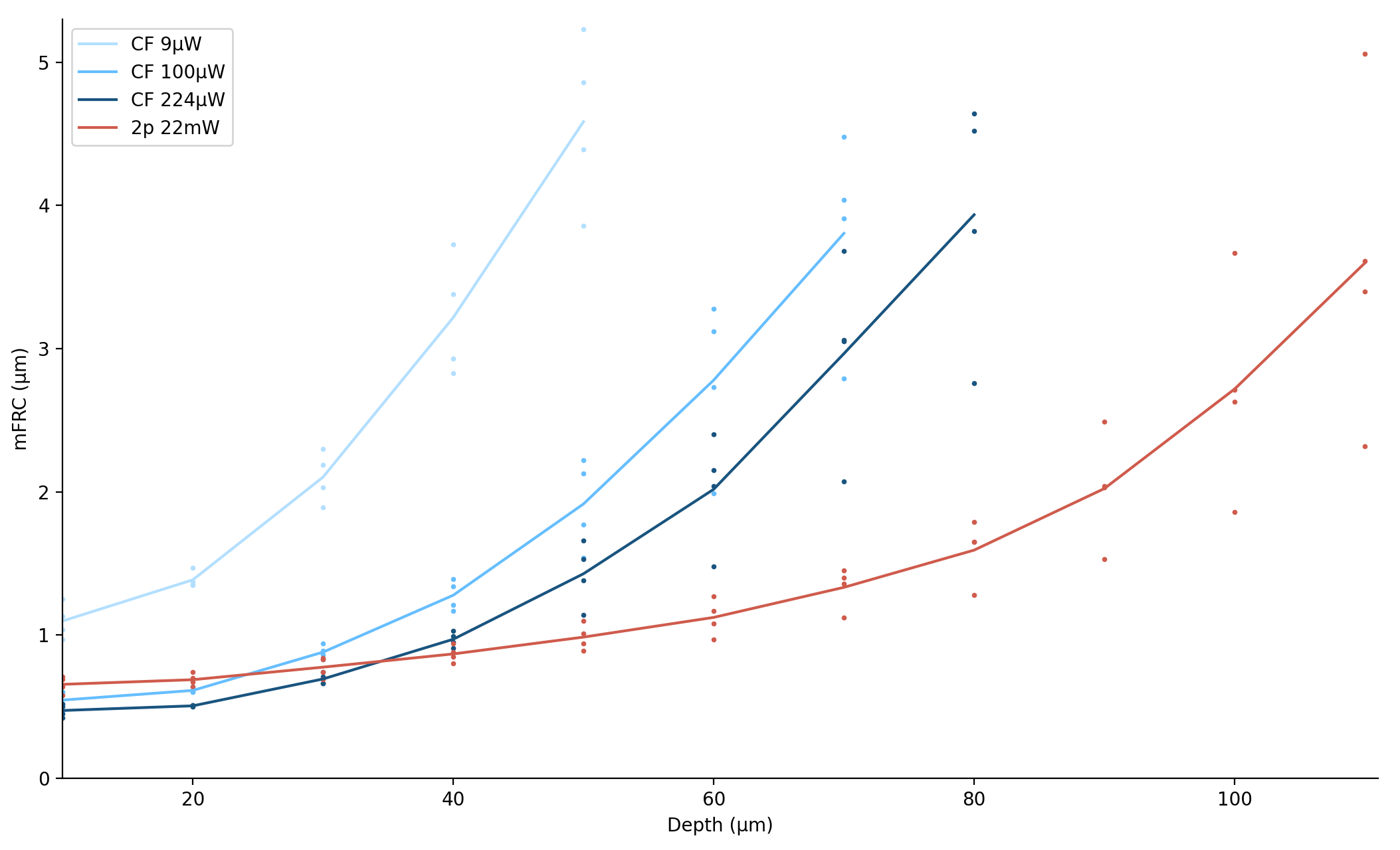}
	\caption{Resolution as a function of depth in the central region of the \textit{Drosophila} brain for 4 different specimens. Values are plotted for depths where $\geq$95\% of tiles contained correlating spatial frequencies for at least half the specimens. Three different confocal power levels (\SI{9}{\micro\watt}, \SI{100}{\micro\watt}, and \SI{224}{\micro\watt}) were compared with two-photon imaging (22 mW) using the mFRC metric. }
	\label{fig:frc_resolution}
\end{figure}

\section*{Results and Discussion}

\subsection*{Theoretical considerations for imaging parameters}

To determine the pixel size required to capture all information within the image, Abbe's diffraction formula was used to estimate the minimum theoretical distance resolvable by the system \cite{rayleigh:1879}:

\begin{equation}
  d = \lambda/(2\text{NA})
\end{equation}

Here, d=349 nm for confocal and 657 nm for 2p imaging. Shannon-Nyquist sampling requires using half these values \cite{shannon:1949}. Therefore, a pixel size of 189 nm (equivalent to full image size of 4096x4096 pixels) was chosen, close to the value required.

Since calculating FRC requires two independent images, initial attempts captured two separate stacks for each modality. However, it was found that occasionally the z-galvo position shifted slightly between stacks, compromising results. To ensure alignment of the two images, `single image' FRC was used instead \cite{koho:2019}, whereby a single image is split into sub-images. To achieve equivalent sampling, stacks were captured with a pixel size of 94.6 nm, equivalent to a pixel size of 189 nm for each sub-image.

\subsection*{Resolution heterogeneity within the same image}

Initial attempts found that using a single FRC calculation for each full image in the stack led to inconsistent and non-monotonic results. FRC colourmaps \cite{zhao:2022_a} revealed that this was due to non-uniformity of resolution within images, particularly the contrast between the central brain and optic lobes, which begin at deeper optical planes (Figure~\ref{fig:algorithm}B) and thus appear as higher resolution areas due to decreased light scattering. Apart from the optic lobes, the effect of differential light scattering on resolution is also apparent at the edges of the central brain, where tissue is thinnest and image quality better, and in the middle of the brain, where tissue is thickest and image quality worst (Figure~\ref{fig:image_matrix}). While the central brain region was analysed here as a whole, depending on requirements it would also be possible to take a more fine-grained approach and analyse only specific structures of interest. Alternatively, the entire brain could be characterised solely as a function of depth using a depthmap-based method. Aside from the effect of scattering, the colourmaps indicated that the outlines of major brain structures, such as the antennal lobes, mushroom body, and central complex, also appeared as high-resolution features, likely due to the presence of sharp `edges' resulting from differing GFP concentrations (Figure~\ref{fig:image_matrix}).

To isolate a particular feature of interest for analysis, simply cropping the image has the disadvantage that only rectangular areas can be analysed, rather than the arbitrary shapes which are common in natural images. An alternative approach of masking out other features risks creating artificially sharp edges in the image, which may distort measurements of resolution. To overcome these issues, the mFRC method was developed based on splitting the image into small tiles. This allows characterisation of resolution in image features of any arbitrary shape, providing results which remain relatively consistent across samples. As noted in the Methods, a `rolling block' approach can be applied similarly if more fine-grained information is required, though this comes at the cost of increased computation.

It is noted that one limitation of this method is that while use of small tiles provides highly localised information, tile size also constrains the lowest spatial frequency which can be correlated. This results in a trade-off between minimising the area to which resolution information is localised and maximising the range of resolutions which can be detected. A tile size of 64$\times$64 pixels was found to provide a reasonable compromise in this regard.

\subsection*{Confocal and two-photon resolution in Drosophila brain}

Two-photon (2p) imaging is widely used in neurobiology, including in \textit{Drosophila} \cite{svoboda:2006}. Use of higher-order excitation suppresses out of focus fluorescence, while longer wavelengths of light are less prone to scattering \cite{helmchen:2005}. Here, the benefits of 2p (22 mW) were apparent for depths approximately \SI{35}{\micro\meter} and greater, with the difference increasing with depth (Figure~\ref{fig:frc_resolution}). For depths less than this, sufficiently-powered confocal appeared to provide slightly sharper images, in line with the theoretical resolution benefits of shorter wavelengths \cite{rayleigh:1879}. Despite the benefits of 2p, however, imaging depth was still limited in comparison to mammalian brain, where imaging up to 1.6 mm has been reported \cite{kobat:2011}. This aligns with a recent study on the optical properties of the fly brain, which attributed its highly scattering nature to extensive light scattering at the air-tissue interface in tracheae \cite{hsu:2019}. At a practical level, though deeper brain structures such as the mushroom body calyx and proto-cerebral bridge could be imaged using 2-photon imaging (Figure~\ref{fig:image_matrix}), higher resolution imaging of these regions may require 3-photon microscopy \cite{hsu:2019}. A further consideration is that 2p causes considerably more photobleaching than confocal, which may be a factor in certain experiments, such as those involving long-term imaging. Additionally, while the results here were based on dissected brains, resolution may vary in fixed samples or \textit{in vivo}.

Confocal laser power was used as an example to demonstrate how improvements in image quality through tuning a single parameter can be quantified using the method described (Figure~\ref{fig:frc_resolution}). In the results, each increase in power led to a measurable improvement in resolution, with the improvement becoming more pronounced with increasing depth. This can be explained by the general principle that, as the number of photons increases, signal increases linearly while noise increases by the square root of the number of photons, leading to a higher signal-to-noise ratio (SNR) \cite{sheppard:2006}. Quantifying resolution in this way thus allows system tuning until requirements for a given experiment are met.

\section*{Conclusion}

The method described showed how any region of arbitrary shape within an image can be characterised in terms of mean resolution using a metric based on Fourier Ring Correlation. This enabled quantification of resolution as a function of depth in 3D features in a way that remains relatively consistent across samples. Using the method applied to the brain of Drosophila, resolution at each depth level of the central brain was measured and the benefits of 2-photon imaging over confocal quantified objectively. Measurement of image quality improvements through tuning a single parameter, laser power, was also demonstrated. It is suggested that this method may be generally useful for other samples in 3D microscopy where resolution is heterogeneous and quantifying image quality at different depths is important.

\section*{Acknowledgements}

NW is grateful for support from the Medical Research Council (MRC). CJR is grateful to the following bodies for support: Engineering and Physical Sciences Research Council (EP/S016538/1, EP/X017842/1, EP/W024969/1); Biotechnology and Biological Sciences Research Council (BB/T011947/1); Wellcome Trust (212490/Z/18/Z); Cancer Research UK (29694, EDDPMA-May22\textbackslash 100059); Royal Society (RGS\textbackslash R2\textbackslash 212305); Chan Zuckerberg Initiative (2020-225443, 2020-225707); Imperial College Excellence Fund for Frontier Research. The Facility for Imaging by Light Microscopy (FILM) at Imperial College London is part-supported by funding from the Wellcome Trust (grant 104931/Z/14/Z) and BBSRC (grant BB/L015129/1). Stocks obtained from the Bloomington Drosophila Stock Center (NIH P40OD018537) were used in this study.

\section*{Disclosures}

The authors declare no conflicts of interest.

\section*{Data Availability Statement}

Data underlying the results presented in this paper are not publicly available at this time but may be obtained from the authors upon reasonable request.

\printbibliography

\end{document}